\documentclass[twocolumn,showpacs,preprintnumbers,amsmath,amssymb,superscriptaddress]{revtex4}
\usepackage{graphicx}% Include figure files
\usepackage{dcolumn}% Align table columns on decimal point
\usepackage{bm}% bold math
\usepackage{hyperref}
\usepackage{latexsym}

\begin{document}

\draft

\title{Analytical solution of a model for complex food webs}

\author{Juan Camacho}
\affiliation{Center for Polymer Studies and Dept. of Physics,
             Boston University, Boston, MA 02215} 
\affiliation{Departament de F\'{\i}sica (F\'{\i}sica Estad\'{\i}stica),
Universitat Aut\`onoma de Barcelona, E-08193 Bellaterra, Spain}

\author{Roger Guimer\`a} 
\affiliation{Center for Polymer Studies and Dept. of Physics, Boston
University, Boston, MA 02215} 
\affiliation{Departament d'Enginyeria Qu\'{\i}mica, Universitat Rovira i
Virgili, 43006 Tarragona , Spain}

\author{Lu\'{\i}s A. Nunes \surname{Amaral}}
%\email{amaral@buphy.bu.edu}
%\homepage{http://polymer.bu.edu/~amaral}
\affiliation{Center for Polymer Studies and Dept. of Physics,
             Boston University, Boston, MA 02215}   
%\author{Juan Camacho$^{1,2}$,
%\thanks{Author to whom correspondence should be addressed}
%Roger Guimer\`a$^{2,3}$ and Lu\'{\i}s A. Nunes Amaral$^{2}$}

%\address{
%	 $^1$ Departament de F\'{\i}sica (F\'{\i}sica Estad\'{\i}stica),
%	 Universitat Aut\`onoma de Barcelona, E-08193 Bellaterra, Spain \\
%         $^2$ Center for Polymer Studies and Department of Physics, 
%        Boston University, MA 02215, USA \\
% 	 $^3$ Departament d'Enginyeria Qu\'{\i}mica, Universitat Rovira i 
%        Virgili, 43006 Tarragona, Catalunya, Spain \\
%}

\begin{abstract}
We investigate numerically and analytically a recently proposed model
for food webs [Nature {\bf 404}, 180 (2000)] in the limit of large web
sizes and sparse interaction matrices. We obtain analytical
expressions for several quantities with ecological interest, in
particular the probability distributions for the number of prey and
the number of predators. We find that these distributions have
fast-decaying exponential and Gaussian tails, respectively. We also
find that our analytical expressions are robust to changes in the
details of the model.
\end{abstract}


\maketitle

%\begin{multicols}{2}

In ecosystems, species are connected through intricate trophic
relationships \cite{Martinez00,McKann98} defining complex
networks \cite{Watts98,Amaral00,Strogatz01,Albert01}, the so-called
food webs.  Understanding the structure and mechanisms underlying the
formation of these complex webs is of great importance in ecology
\cite{Cohen90,Polis91,Rejmanek,Briand}. In particular, food web
structure provides insights into the behavior of ecosystems under
perturbations such as the introduction of new species or the
extinction of existing species. The nonlinear response of the elements
composing the network leads to possibly catastrophic effects for even
small perturbations \cite{weak}.

%Thus, the practical implications of
%food web structure are of great importance for situations such as crop
%selection, management of fishing stocks, preservation of threatened
%ecosystems and maintenance of biodiversity \cite{biodiversity}.
%Theoretical analyses of food webs may then supply valuable practical
%information for the management of ecological resources.

Recently, Williams and Martinez have proposed an elegant model of food
webs ---the ``niche'' model--- that just with a few ingredients
successfully predicts key structural properties of the most
comprehensive food webs in the literature \cite{Martinez00}.
Numerical simulations of the niche model predict values for many
quantities typically used to characterize empirical food webs that are
in agreement with measured values for seven webs in a variety of
environments, including freshwater habitats, marine-freshwater
interfaces and terrestrial environments.

%%%%%%%%%%%%%%%%%%%%%%%%%%%%%%%%%%%%%%%%%%% RESULTS

We investigate the niche model from a theoretical perspective. We
study analytically and numerically the behavior of key quantities for
sparse food webs, i.e., webs with $L \ll S^2$, where $L$ is the number
of trophic interactions between species and $S$ is the number of
species in the web.  This is the limit of interest in ecology because
(i) for most food webs reported in the literature the directed
connectance, defined as $C=L/S^2$ takes small values, and (ii) it
corresponds to the limit of large web sizes $S$
\cite{Rejmanek,Briand}.
%We derive here analytical expressions that
%enable us to predict the behavior of the model beyond its specific
%outputs for a limited choice of parameter values.
We calculate the probability distributions of number of prey and of
number of predators and find that for $C \ll 1$ they depend only on
one parameter of the model---the average number $z$ of trophic links
in the network.  These distributions give valuable information about
the structure of the network \cite{attack} and enable us to calculate
other interesting quantities such as the fraction of ``top,''
``intermediate,'' and ``basal'' species, and the standard deviation of
the ``vulnerability'' and ``generality'' of the species in the food
web \cite{Martinez00}.  Our results provide compact patterns that
describe the structure of the food webs generated by the niche model.
These patterns could not have been predicted from the numerical
simulations reported in Ref.~\cite{Martinez00} and may be of practical
and fundamental importance for the study of empirical food
webs. Moreover, we test our analytical predictions with empirical food
webs and find agreement.

%%%%%%%%%%%%%%%%%%%%%%%%%%%%%%%%%%%%%%%%% THE MODEL

Next, we define the model.  Consider an ecosystem with $S$ species and $L$
trophic interactions between these species.  These species and interactions
define a network with $S$ nodes and $L$ directed links. In the model, one
first randomly assigns $S$ species to ``trophic niches'' $n_i$ mapped into
the interval [0,1].  For convenience, we will assume that the species are
ordered according to their niche number, i.e., $n_1 < n_2 < ... < n_S$.

A species $i$ is characterized by its niche parameter $n_i$ and by its
list of prey.  Prey are chosen for all species according to the
following rule: a species $i$ preys on all species $j$ with niche
parameters $n_j$ inside a segment of length $r_i$ centered in a
position chosen randomly inside the interval $[r_i/2,n_i]$, with $r_i
= x n_i$ and $0 \le x \le 1$ a random variable with probability
density function
\begin{equation}
p_{x}(x) = b \left(1-x\right) ^{\left( b-1\right)}\,.
\end{equation}
The values of parameters $b$ and $S$ determine the average
connectivity $z \equiv 2 L / S$ of the food web and the directed
connectance $C = L / S^{2} $\cite{Martinez00,note1}.  One can also
express the average number of prey per species as $S\overline{r}$,
where the bar indicates an average over an ensemble of food
webs. It then follows that the connectivity is $z=2S\overline{r}$, the
number of directed links is $L=S^{2}\overline{r}$, and the connectance
is $C = \overline{r}$. One can also obtain these expressions in terms
of $b$ using the equality $\overline{r}=\overline{x}/2=1 / [2(1+b)]$.

In the  niche model, isolated species---that is, species with no prey or
predators---are eliminated and species with the same list of prey and
predators ---that is, trophically-identical species--- are ``merged''
\cite{note11}.

%%%%%%%%%%%%%%%%%%%%%%%%%%%%%%%%%%%%%% PREY

First, we address the statistics of the number of prey. For large
$S$, the number of prey of a species $i$ is $k_i = S r_i$, so that the
probability distribution $p_{\rm prey}$ is given directly by the
distribution of $r$.  Specifically, $p_{\rm prey}(k) = p(r) / S$.

The cumulative probability $P(r'>r) = \int _{r}^{1}dr'p(r')$ is the
area of the region $R$ of the \( n-x\textrm{ } \)diagram bounded by
lines $x = 1$, $n = 1$, and the hyperbole $r = nx$
\begin{equation}
\label{cump}
P(r'>r) = \int_r^1\, dx\, \int_{r/x}^1\, dn\, p_{n}(n)\, p_{x}(x) \,,
\end{equation}
where $p_{n}(n) = 1$ is the probability density function of $n$.  The
integration of (\ref{cump}) gives rise to a function involving
hypergeometric functions \cite{numrec}.  To obtain a more
``physical'' solution, one can derivate (\ref{cump}) twice to obtain
\begin{equation}
\label{expon}
\frac{dp(r)}{dr}=-\frac{p_{x}(r)}{r} \,. 
\end{equation}
In the limit $C \ll 1$, one has $b \gg 1$ (see \cite{note11}), so that
$p_{x}(x) \simeq be^{-bx}$, and the term in the right-hand side
vanishes exponentially, indicating the $p(r)$ and $P(r'>r)$ have
exponentially decaying tails \cite{note15}.

\begin{figure}[t!]
\centerline{\includegraphics*[width=0.75\columnwidth]{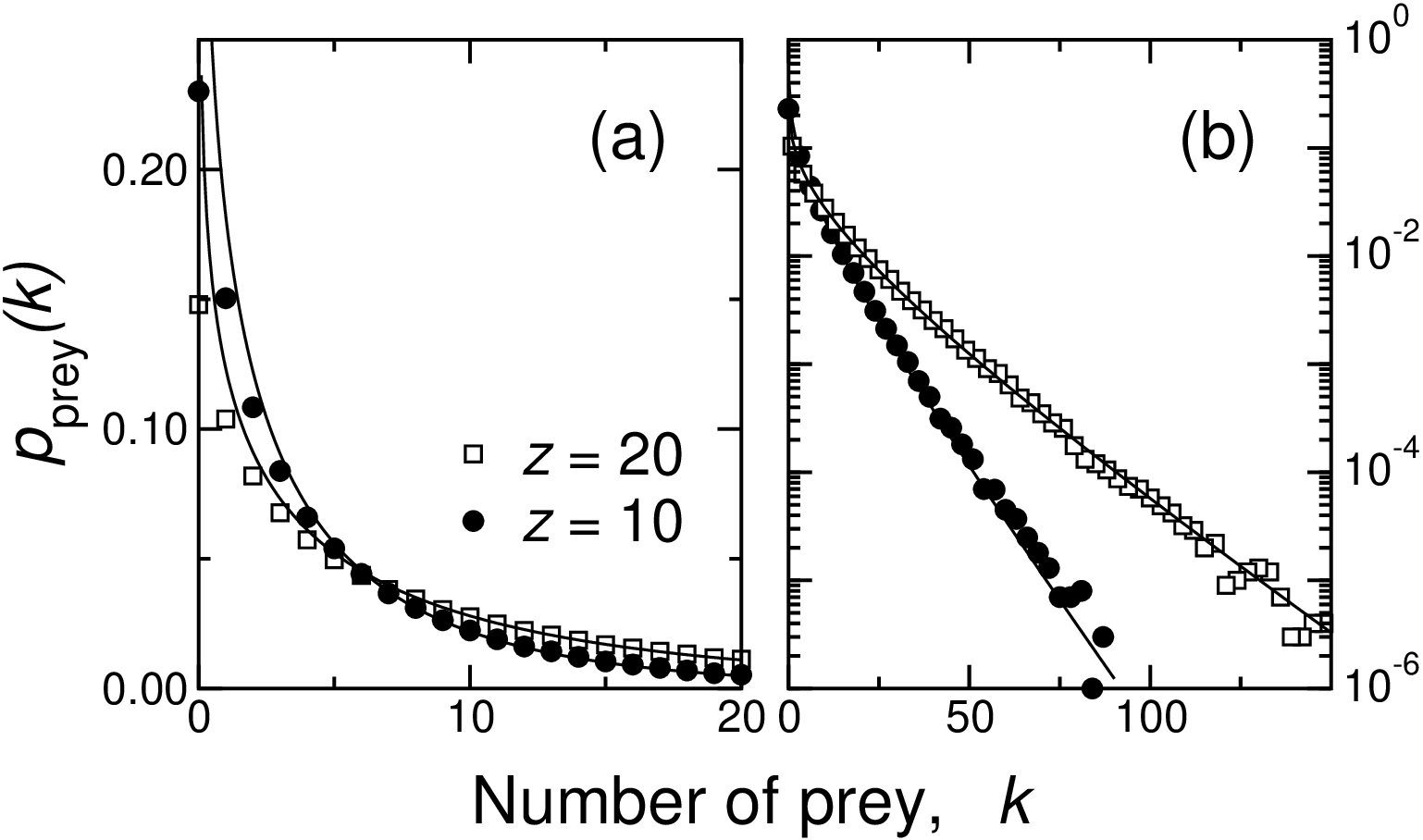}}
\centerline{\includegraphics*[width=0.75\columnwidth]{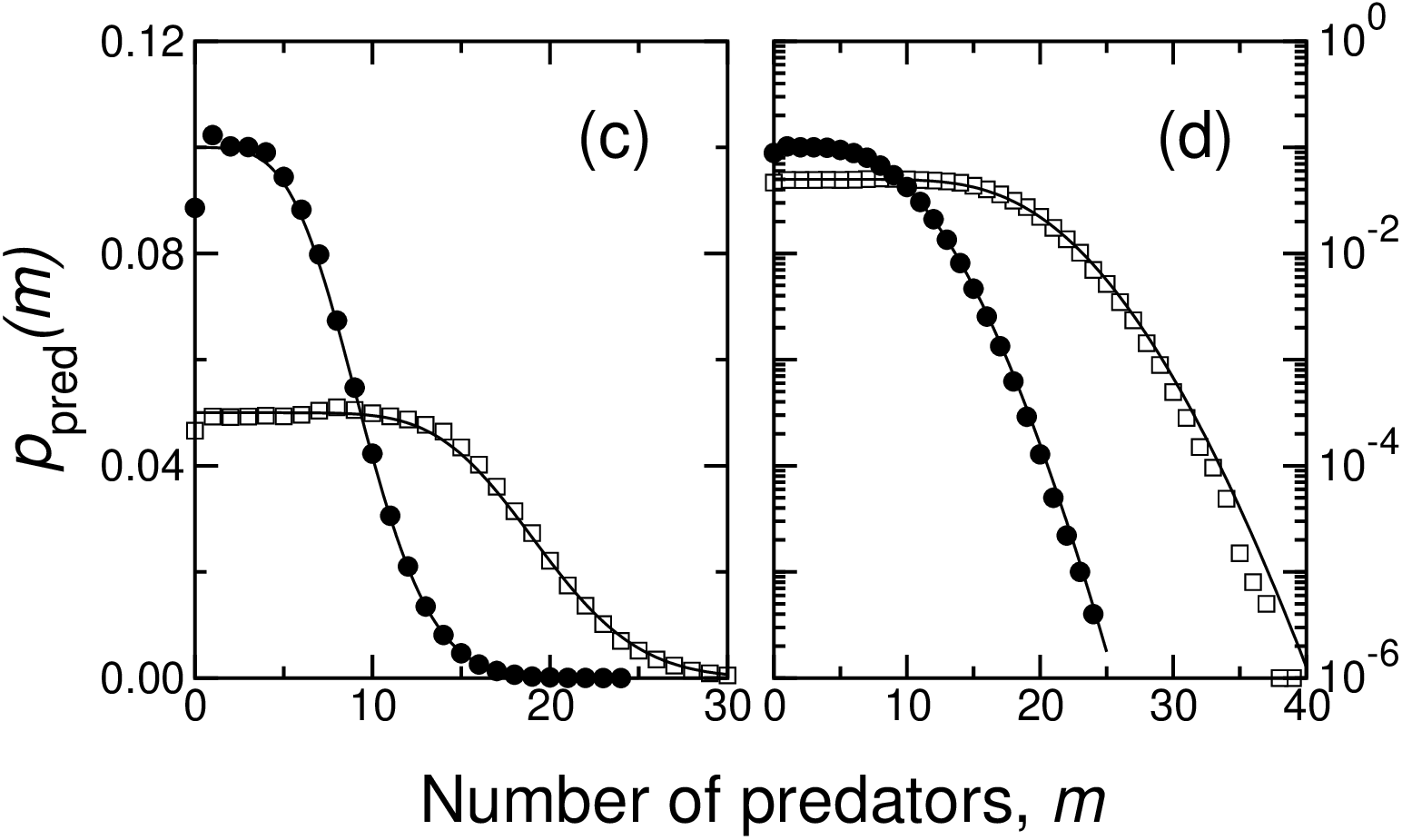}}
\centerline{\includegraphics*[width=0.72\columnwidth]{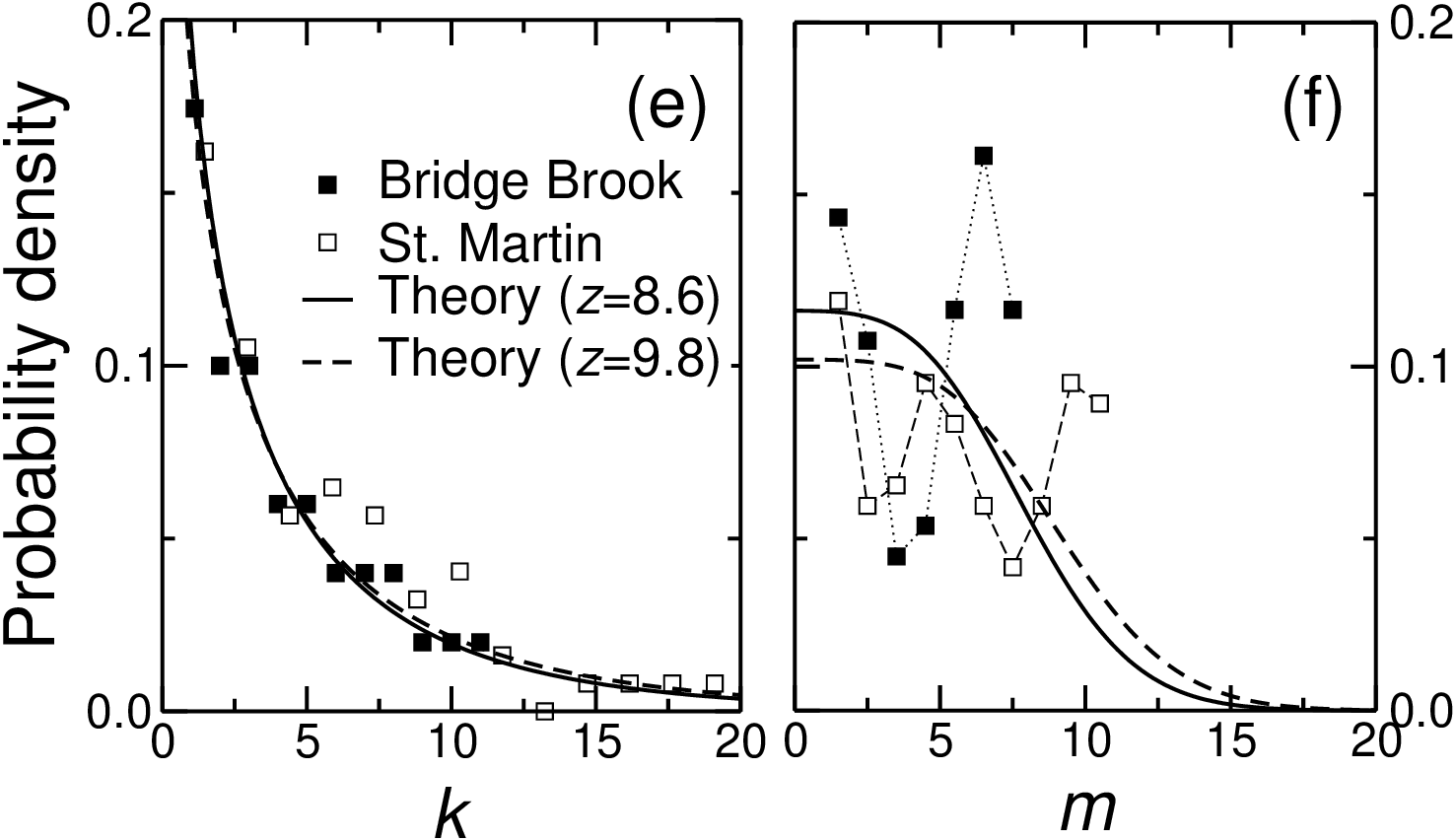}}
%\vspace{0.1in}
\caption{ (a) Linear and (b) log-linear plots of the distribution of
number of prey for 1000 simulations of food webs with $S=1000$.  We
show results for $z=10, 20$ and the corresponding theoretical
predictions.  As expected, we find an exponential decay of the
distributions. (c) Linear and (d) log-linear plots of the distribution
of number of predators for the same food webs as in (a) and (b).  As
predicted, we find a regime where the distribution is uniform followed
by a Gaussian decay. We test our analytical predictions with empirical
data \protect\cite{Martinez00} for (e) $p_{\rm prey}(k)$ and (f)
$p_{\rm pred}(m)$ for Bridge Brook (solid line) and St. Martin (broken
line).
\label{fig1}}
\end{figure}

To obtain a simpler analytical solution for $p(r)$ than given by the
hypergeometric functions, we approximate $p_{x}$ in the entire
$x$-range by an exponential.  We expect the results to be the same for
$\overline{x} \ll 1$ \cite{note11} because $p_{x}$ takes non-vanishing
values only for small $x$. Under this approximation, the integration
of (\ref{expon}) yields
\begin{equation}
\label{presas}
p(r)=bE_{1}(br) \,,
\end{equation}
where $E_{1}(x) = \int ^{\infty }_{x} dt~t^{-1} \exp(-t)$ is the
exponential-integral function \cite{numrec}.  The probability
distribution $p_{\rm prey}(k)$ is obtained from (\ref{presas}) making
the substitutions $r=k/S$ and $b = S/z$, yielding
\begin{equation}
\label{prey}
p_{\rm prey}(k)=(1/z)\, E_{1}(k/z) \,.
\end{equation}
We compare in Figs.~\ref{fig1}(a) and (b) the predictions of
(\ref{presas}) with numerical simulations.  We find close agreement
between our analytical expression and the numerical results.  In
particular, they show an exponential decay for large $k$. The
deviations observed for small values of $k$ are due to the fact that
$k_{j}=S r_{j}$ is an average value implying that it is a good
approximation only when the fluctuations of $k_{j}$ are small, which
is no longer true for small $k$.

%%%%%%%%%%%%%%%%%%%%%%%%%%%%%%%%%%%%%% PREDATORS

Next, we address the statistics of the number of predators. Note that
for $\overline{r} \ll 1$ \cite{note11}, the predators of species $i$
have, to first approximation, niche values $n_j > n_{i}$ and that the
segment $r_j$ is placed with equal probability in the interval $[0,
n_j]$. Therefore, the probability for a species $j$ to prey on $i$ is
$r_{j} / n_{j}= x_j n_j / n_j = x_j$, implying that the average
probability for the species with $n_j > n_i$ to prey on species $i$ is
$\overline{x}$.

If we assume that $S \gg 1$, the number of predators of $i$ is the
result of $S-i$ independent ``coin-throws'' with probability
$\overline{x}$ of being a predator and probability $1-\overline{x}$ of
not being a predator, implying that the probability of species $i$
having $m$ predators is given by the binomial distribution. It then
follows that the distribution of number of predators for a general
species is the average over the different binomials
\begin{equation}
\label{predators}
p_{\rm pred}(m)=\frac{1}{S}\sum _{i=1}^{S-m}{{S-i}\choose m}
\overline{x}^{m} (1-\overline{x})^{S-m-i}
\end{equation}
In the limit of interest, $S \gg 1$, $\overline{x} \ll 1$,
and $S \overline{x} = z$, one can approximate the binomial
distribution by a Poisson and the sum by an integral
\begin{equation}
\label{pred}
p_{\rm pred}(m)=\frac{1}{z}\int _{0}^{z}dt\, \frac{t^{m}e^{-t}}{m!}=
\frac{1}{z}~\gamma (m+1,z),
\end{equation}
where $\gamma$ is the ``incomplete gamma function''
\cite{numrec,note3}.  For $m < z/2$, the function $\gamma$ is
approximately constant, while it decays with a Gaussian tail for $m
\approx z$.  In Fig.~\ref{fig1}(c) and (d), we compare the predictions
of (\ref{pred}) with numerical solutions and find good agreement.

In Figs.~\ref{fig1}e,f, we compare our analytical predictions,
Eqs.~(\ref{prey})--(\ref{pred}), with data from recent food webs:
Bridge Brook ($S=25, z=8.6$) and St.\ Martin Island ($S=42,
z=9.8$). We find that the distributions of number of prey is well
approximated by the data and that the distributions of number of
predators are ``noisy'' but still show the expected cutoff for $m
\approx z$ and is approximately constant for $m < z$ as predicted by
Eq.~(\ref{pred}).  This agreement is remarkable since the webs
analyzed are quite small so one might not expect the theoretical
expressions to hold.
%%%%%%%%%%%%%%%%%%%%%%%%%%%%%%%%%%%%%%% TOP AND BASAL

Next, we evaluate the fraction of top $T$, intermediate $I$ and basal
$B$ species. As the names indicate, top species have no predators and
basal species have no prey, while intermediate species are those with
both prey and predators. The fraction of intermediate species is just
$I = 1 - (T + B)$. The fraction $T$ of top species is, by definition,
\vspace{-3mm}
\begin{equation}
\label{top}
T \equiv p_{\rm pred}(0) =\frac{1-\exp(-z)}{z}\,.
\end{equation}
Note that a similar result is obtained if one calculates the sum
(\ref{predators}) for $m=0$. Since typically $5 < z < 20$, Eq.~(\ref{top})
can be approximated simply as $T=1/z$.

To calculate the fraction $B$ of basal species, we note that a species
has no prey only if its range $r$ falls in a region with no species
\cite{note30}. In the limit of large S, the probability density for
finding an empty interval of length $\delta$ is $S e^{-S\delta}$, as
predicted by the canonical distribution \cite{pathr}. Thus, the
probability of finding a species-free segment of length larger than
$r$ is $e^{-Sr}$, which gives the probability for a species of range
$r$ not to prey on other species.  Using (\ref{presas}), it follows
that the average probability is:
\vspace{-3mm}
\begin{equation}
\label{basal}
B = \int _{0}^{1}dr\, e^{-Sr}p(r) = \frac{\ln(1+z)}{z}.
\end{equation}

\begin{figure}[t!]
\centerline{\includegraphics*[width=0.75\columnwidth]{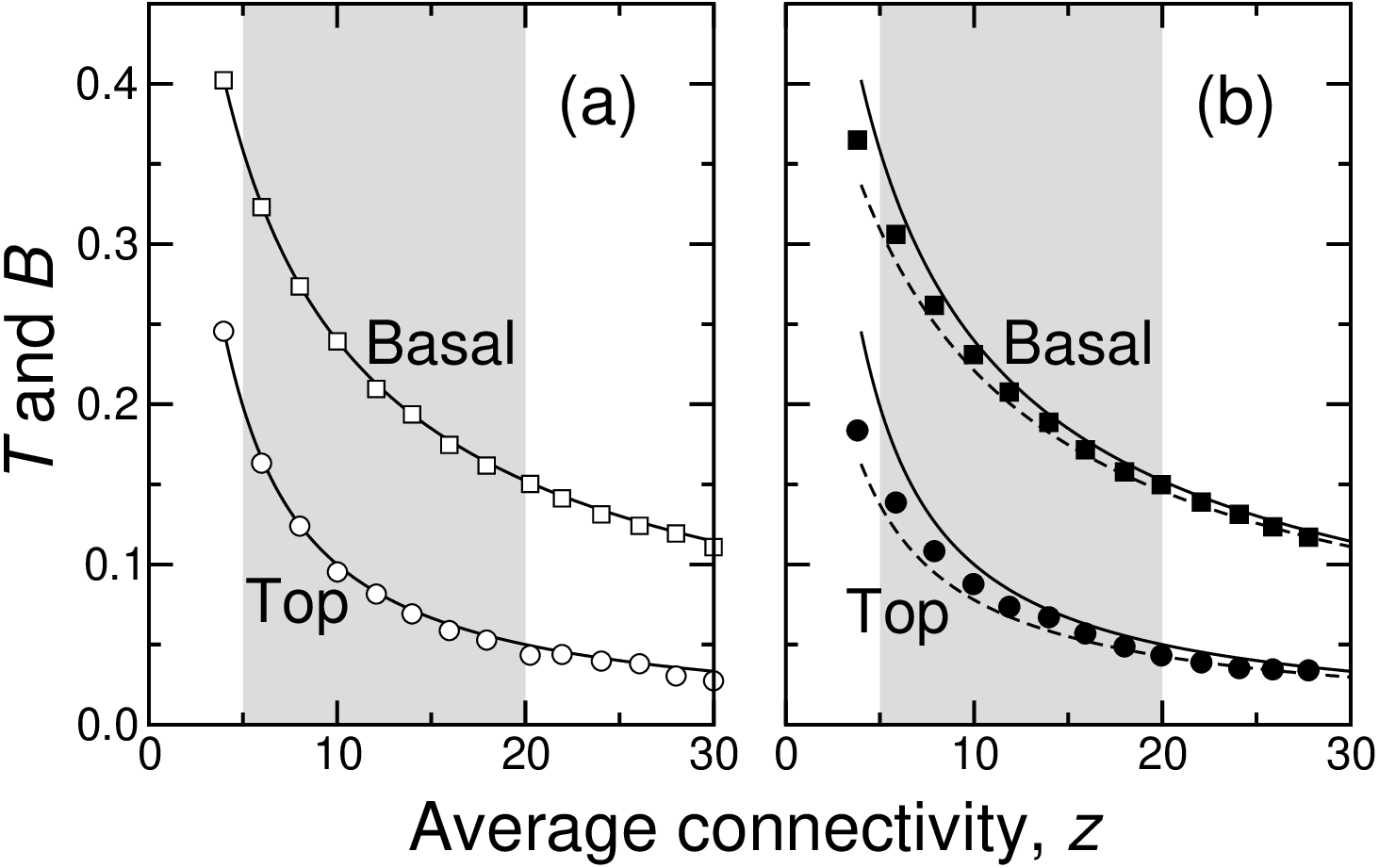}}
%\vspace{0.1in}
\caption{Fraction of top and basal species as a function of the
average connectivity $z$.  The shaded region corresponds to the
interval of $z$ typically observed in empirical food webs. (a)
Comparison of the results of 100 simulations of food webs with
$S=1000$---for which isolated species where {\it not\/} removed---with
our theoretical predictions
Eqs.~(\protect\ref{top})--(\protect\ref{basal}). Note the good
agreement between the analytical expressions and the numerical
results. (b) Comparison of the results of 100 simulations of food webs
with $S=1000$---for which isolated species were removed---with our
theoretical predictions
Eqs.~(\protect\ref{top})--(\protect\ref{tb1}). Note that the
theoretical predictions provide narrow bounds for the numerical
results.
\label{fig2}}
\end{figure}

In the model \cite{Martinez00}, isolated species are eliminated, so
they are not counted towards top or basal species.  To correct the
estimates (\ref{top})--(\ref{basal}) for this effect, we remove the
isolated species.  We estimate the number of isolated species to first
order by assuming that having no prey is statistically independent of
having no predators, implying that the fraction of isolated species is
just the product of the fractions of top and basal species.  This
assumption does not take into account the possibility that a species
with no prey is likely to have a low niche value $n$ and hence it has
a high probability to have predators.  Nonetheless, this simple
approximation provides an upper bound for the number of isolated
species which leads to a lower bound on $T$ and $B$,
\begin{equation}
\label{tb1}
T'=\frac{T-TB}{1-TB}\,, \quad\quad\quad B'=\frac{B-TB}{1-TB}
\end{equation}
In Fig.~\ref{fig2}, we compare our analytical predictions for the fraction of
top and basal species with numerical simulations of the model. As expected,
Eqs.~(\ref{top})--(\ref{tb1}) provide bounds for the numerical results.

%%%%%%%%%%%%%%%%%%%%%%%%%%%%%%%%%%%% VUL AND GEN

Finally, we calculate the standard deviations of the vulnerability and
generality of the species in food webs generated according to the
model.  The vulnerability of a prey is defined as its number $m$ of
predators, and the generality of a predator as its number $k$ of prey.
Following Ref.\ \cite{Martinez00}, we define the normalized standard
deviations of the vulnerability as $\sigma_V^2=
\overline{m^{2}}/\overline{m}^{2}-1$ and of the generality as
$\sigma_G^2= \overline{k^{2}}/\overline{k}^{2}-1$. By definition, one
has $\overline{m}=\overline{k}=z/2$ for both cases.

\begin{figure}[t!]
\centerline{\includegraphics*[width=0.75\columnwidth]{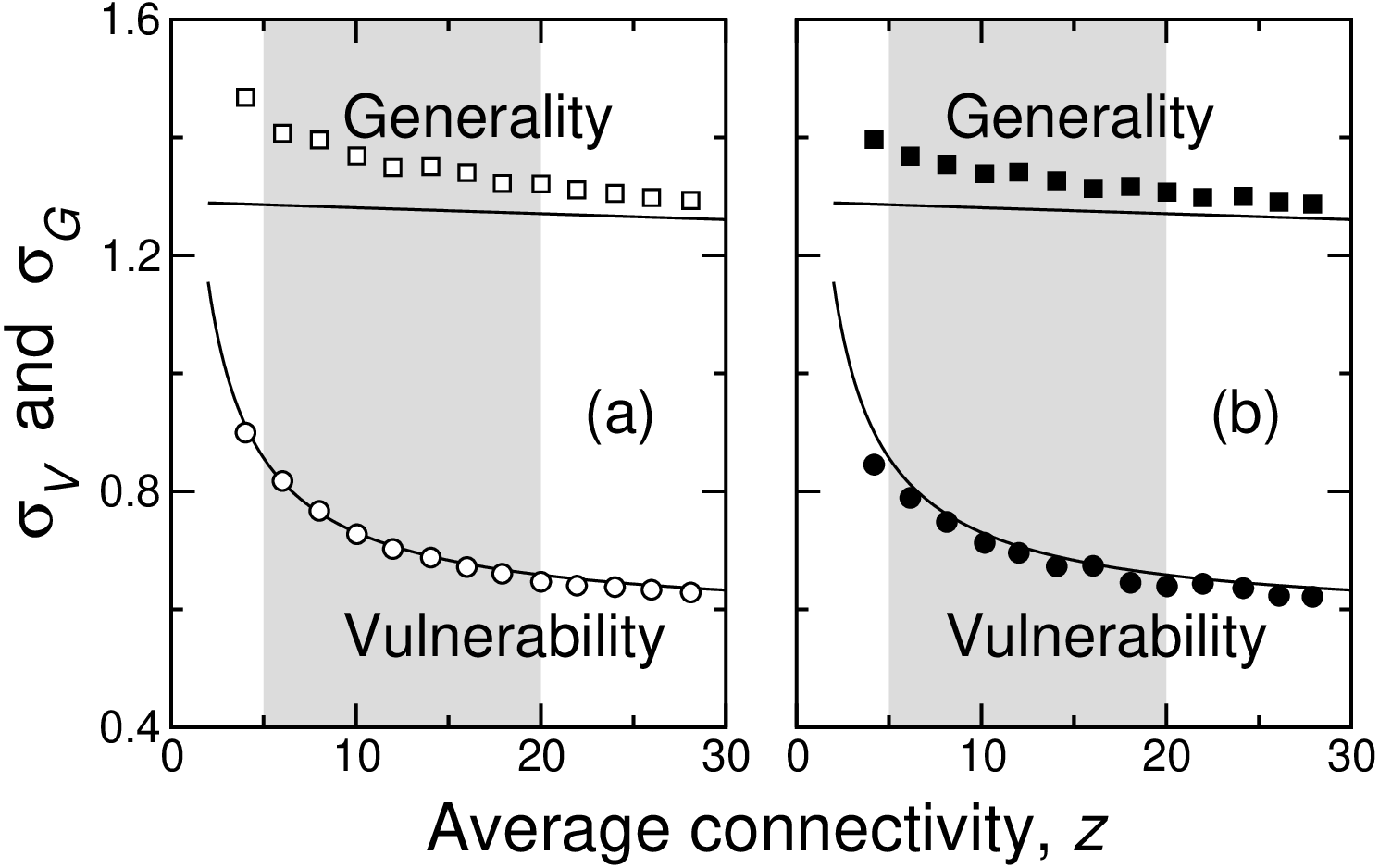}}
%\vspace{0.1in}
\caption{Normalized standard deviations of generality and
vulnerability as a function of the average connectivity $z$.  The
shaded region corresponds to the interval of $z$ typically observed in
empirical food webs. (a) Comparison of the results of 100 simulations
of food webs with $S=1000$---for which isolated species where {\it
not\/} removed---with our theoretical predictions
Eqs.~(\protect\ref{deltaV})--(\protect\ref{deltaG}).  (b) Comparison
of the results of 100 simulations of food webs with $S=1000$---for
which isolated species where removed---with our theoretical
predictions Eqs.~(\protect\ref{deltaV})--(\protect\ref{deltaG}).  Note
that as for Fig.~\protect\ref{fig2}, removing isolated species leads
to slightly less good agreement with the simulation results for
$\sigma_V$.  However, the removal of isolated species does not appear
to be a factor in the deviations found for $\sigma_G$.  The reason why
$\sigma_G$ underestimates the simulation results at small $z$ values
relates to the fact that $k_j = S r_j$ is a good approximation only
when the fluctuations of $k_j$ are small, which is no longer true for
small $k$.
\label{fig3}}
\end{figure}

To evaluate $\sigma_V$, we first calculate
$\overline{m^{2}}$. Equation~(\ref{pred}) yields $\overline{m^{2}}=
%\frac{1}{z} \int_{0}^{z} (x^{2}+x) 
z^2 / 3 + z / 2$, so that
\begin{equation}
\sigma_{V}=\sqrt{\frac{1}{3}+\frac{2}{z}}.
\label{deltaV}
\end{equation}
We next calculate $\sigma_G$, for any value of $C$, by direct
evaluation of $\overline{k^{2}}$. If $S \gg 1$, the number of prey of
a species having a range $r$ is $k=S r$, and we find that
$\overline{k^{2}} / \overline{k}^{2} = \overline{r^{2}} /
\overline{r}^{2} = 8 (b+1) / [3(b+2)],$ implying that
\begin{equation}
\sigma_{G}=\sqrt{\frac{8}{3} \frac{1}{1+2 C}-1}.
\label{deltaG}
\end{equation}
For $C \ll 1$, $\sigma_G$ becomes a constant with value $\sqrt{5/3}$,
a result that can also be obtained from (\ref{prey}).  We show in
Fig.~\ref{fig3} the results for our analytical expressions
(\ref{deltaV}) and (\ref{deltaG}) and compare them with results from
numerical simulations of the niche model.

%%%%%%%%%%%%%%%%%%%%%%%%%%%%%%%%%%%%%%%
%%%%%%%%%%%%%%% Conclusion

We have also studied the robustness of our predictions to changes in
the particular formulation of the {\it details} of the model.  The
nature of approximations used in the derivations of the expressions
for the distributions of the number of prey and predators,
Eqs.~(5)-(7), allow us to conclude that: (i) the distribution of the
number of predators does not depend on the specific form of
$p(x)$. The only requirement is that the connectance
$C=\overline{x}/2$ tends to zero under some limit, so that $z=S C$
remains finite when $S$ tends to infinity; and (ii) the distribution
of the number of prey depends on the functional form of $p(x)$, but
Eq.~(7) will still be is obtained for all $p(x)$ decaying
exponentially as $C$ tends to zero. Thus, it appears that our findings
are robust under quite general conditions, a result that is not
possible to obtain without an analytic treatment of the problem.

Our results are also of interest for a number of other reasons.
First, we demonstrate for the first time that the distributions of
number of prey and number of predators have different functional
forms.  Second, we show that both distributions have characteristic
scales, i.e., both distributions have well defined means and standard
deviations as $S$ increases to infinity.  Third, we find that the
functional forms of the distributions of number of prey and number of
predators depend only on the average connectivity $z$, and agree with
empirical data. This result is rather surprising in face of the
complexity of the empirical and model food webs. Finally, we show that
other quantities of biological interest also depend exclusively $z$.

\begin{acknowledgments}
We thank A. Arenas, J. Bafaluy, M. Barth\'el\'emy,
A. D\'{\i}az-Guilera, G. Franzese, F. Giralt, E. LaNave, S. Mossa,
A. Scala, H. E. Stanley and especially N. D. Martinez and
R. J. Williams for stimulating discussions and helpful suggestions.
JC and RG thank the Generalitat de Catalunya and the Spanish CICYT
(PB96-0168, PB96-1011 and BFM2000-0351-C03-01) for support. LANA
thanks NIH/NCRR (P41 RR13622) for support.
\end{acknowledgments}

%\vspace*{-0.7cm}

%%%%%%%%%%%%%%%%%%%%%%%%%%%%%%%%%%%%%%%%%%%%%%%%%%%%%%%%%%%%%%%%%%%%%%%
%%%%%%%%%%%%%%%%%%%%%%%%%%%%%%%%%%%%%%%%%%%%%%%%%%%%%%%%%%%%%%%%%%%%%%%
%%%%%%%%%%%%%%%%%%%%%%%%%%%%%%%%%%%%%%%%%%%%%%%%%%%%%%%%%%%%%%%%%%%%%%%
%%%%%%%%%%%%%%%%%%%%%%%%%%%%%%%%%%%%%%%%% REFERENCES

%\end{multicols}


\begin{references}
\bibitem{note1} The ``highest-quality'' food webs reported in the
literature have 50--100 species with $10 < z < 20$ and connectance of
0.1--0.3 \protect\cite{Martinez00}. Empirical studies suggest that $z$
is either a constant \protect\cite{Rejmanek} or grows as $S^{0.4}$
\protect\cite{Briand}, implying that $C$ tends to zero for large food
webs.
\bibitem{note11} Note that the limit $C \ll 1$ is equivalent to
$\overline{x}$ and $\overline{r} \ll 1$ and $b \gg 1$. For
$\overline{r} \ll 1$ the probability that two species are trophically
identical is very small.  This suggests that taxonomic and trophic
classifications of species lead to similar results \cite{Martinez00}.
\bibitem{note15} The exponential decay is found when $\overline{r} \ll
r \ll 1$. When $r$ approaches one, the decay is much faster and $p(r)$
becomes identically zero.
\bibitem{note3} There are two incomplete gamma functions defined in
the literature, $\gamma(\alpha,x)=\int_{0}^{x} e^{-t}
t^{\alpha-1}/\Gamma(\alpha)$ and $\Gamma(\alpha,x)=\int_{x}^{\infty}
e^{-t} t^{\alpha-1}/\Gamma(\alpha)$. In some cases they are defined
without the $\Gamma(\alpha)$ in the denominator.
\bibitem{note30} The definition of the fraction $B$ of basal species,
$p_{\rm prey}(0)$, leads, in the continuous analysis, to a wrong
approximation because the exponential-integral function $E_{1}(x)$
diverges as $x$ approaches zero.
\end{references}
\end{document}